\documentclass[aps,prl,twocolumn,showpacs,groupedaddress]{revtex4-1}
\pdfoutput=1
\usepackage{dcolumn}   
\usepackage{bm}        
\usepackage{amssymb}   
\usepackage{amsfonts}
\usepackage{verbatim}
\usepackage[pdftex]{graphicx}  
\usepackage{amsmath}
\usepackage{amsfonts}
\usepackage{ulem}
\usepackage[pdftex]{color}

\newcommand{\ket}[1]{|#1\rangle}

\def\lsim{\mathrel{\rlap{\lower4pt\hbox{\hskip1pt$\sim$}}
    \raise1pt\hbox{$<$}}}                
\def\gsim{\mathrel{\rlap{\lower4pt\hbox{\hskip1pt$\sim$}}
    \raise1pt\hbox{$>$}}}                

\begin{document}
\normalem

\title{Coherent Josephson phase qubit with a single crystal silicon capacitor}
\author{U. Patel}
\author{Y. Gao}
\author{D. Hover}
\author{G. J. Ribeill}
\author{S. Sendelbach}
\author{R. McDermott}
\email[Electronic address: ]{rfmcdermott@wisc.edu}

\affiliation{Department of Physics, University of Wisconsin, Madison, Wisconsin 53706, USA}

\date{\today}

\begin{abstract}
We have incorporated a single crystal silicon shunt capacitor into a Josephson phase qubit. The capacitor is derived from a commercial silicon-on-insulator wafer. Bosch reactive ion etching is used to create a suspended silicon membrane; subsequent metallization on both sides is used to form the capacitor. The superior dielectric loss of the crystalline silicon leads to a significant increase in qubit energy relaxation times. $T_1$ times up to 1.6~$\mu$s were measured, more than a factor of two greater than those seen in amorphous phase qubits. The design is readily scalable to larger integrated circuits incorporating multiple qubits and resonators.
\end{abstract}

\pacs{85.25.Cp,77.22.Gm,74.50.+r}
\maketitle

The Josephson phase qubit is an attractive candidate for scalable quantum information processing in the solid state \cite{Clarke08,Martinis02,Lucero12,Neeley10,Ansmann09}. This qubit has achieved several important milestones, including realization of high-fidelity entangling gates in two and three qubit circuits  \cite{Neeley10}, violation of a Bell's inequality \cite{Ansmann09}, and full characterization of highly non-classical states in linear microwave resonators \cite{Hofheinz09}. However, qubit performance is limited by relatively short coherence times, under 1~$\mu$s. It has been shown that the dominant energy relaxation mechanism is dielectric loss induced by a continuum of low-energy defect states in the amorphous thin films of the circuit \cite{Martinis05}. In the most common approach to qubit realization, a $\sim$ 1~$\mu$m$^2$ Josephson  junction is shunted by an external thin film capacitor of order 1~pF \cite{Steffen06}; in this case, qubit $T_1$ is determined solely by the loss tangent of the bulk capacitor dielectric: $T_{1}={1}/{\omega_{10}\tan\delta}$, where ${\omega_{10}/2\pi}$ is the qubit frequency. There have been efforts to develop amorphous dielectrics with improved intrinsic (low-temperature, low power) loss for superconducting qubit applications \cite{OConnell08}. An alternative approach is to incorporate crystalline, defect-free dielectrics into the qubit circuit. Molecular beam epitaxy techniques have been used to grow single-crystal Al$_{2}$O$_{3}$ tunnel barriers on crystalline Re underlayers, and phase qubit circuits incorporating these barriers demonstrate a significant reduction in the density of spurious microwave resonances \cite{Oh06,Kline09}. Moreover, there has been progress in the incorporation of epitaxial Josephson junctions into transmon qubits, although $T_1$ times were under 1~$\mu$s \cite{Nakamura11,Weides11}. Other work involves the development of grown bulk crystalline Al$_{2}$O$_{3}$ for phase qubit shunt capacitors \cite{Cho}. However, the robust, repeatable growth of crystalline dielectrics for superconducting qubit circuits remains a daunting challenge.

\begin{figure}[!t]
\includegraphics[width=.48\textwidth]{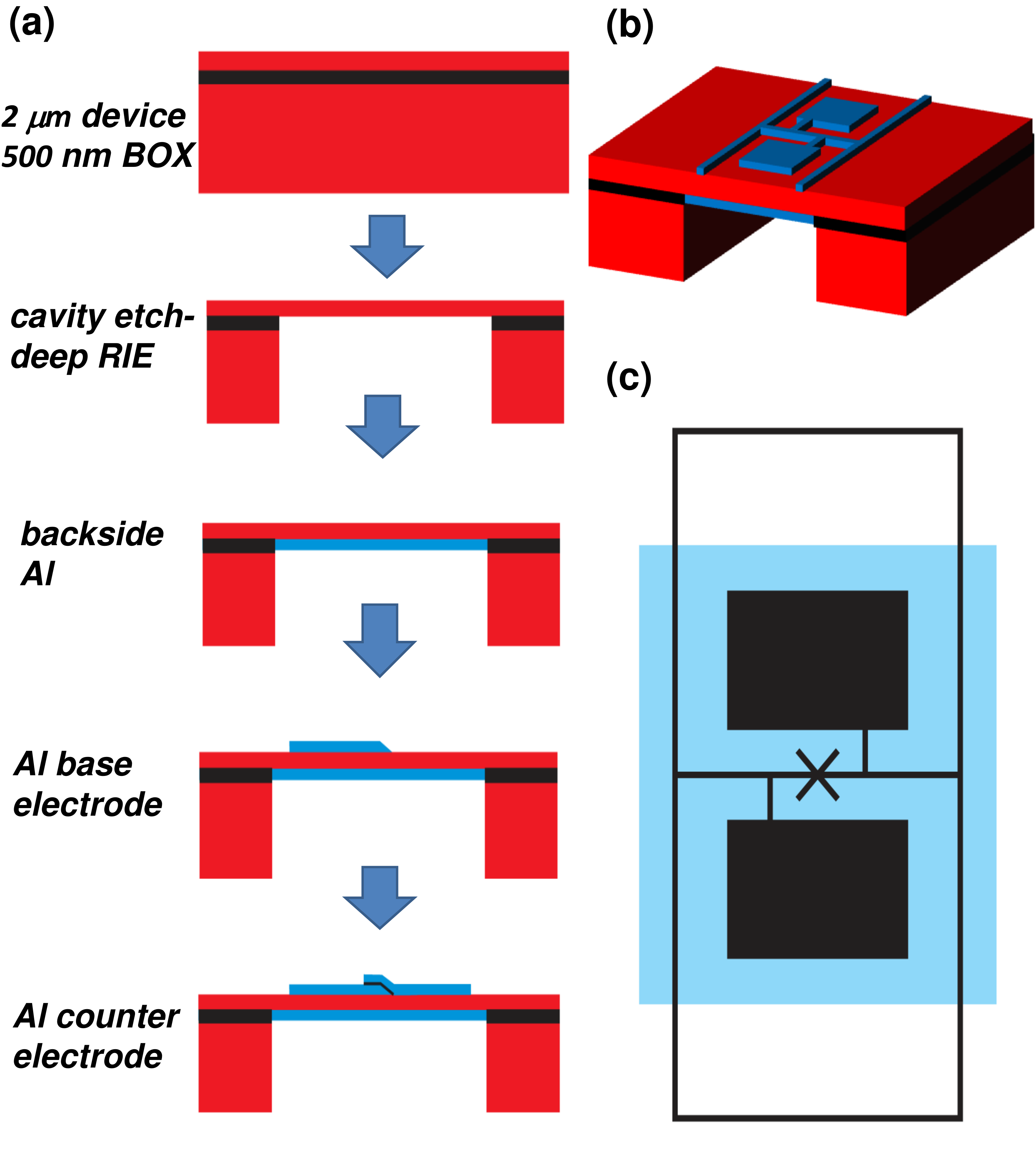}
\vspace*{-0.0in} \caption{Superconducting Josephson phase qubit with crystalline silicon shunt capacitor. (a) Fabrication process flow. The cavity etch is followed by Al deposition on the front and back sides. The front side Al is patterned into a base wiring layer; subsequent ion milling, oxidation, and counterelectrode deposition and etch steps form the Al-AlO$_x$-Al Josephson junction. (b) 3D view of the layer stackup for the completed device. (c) Phase qubit circuit layout. The qubit loop inductance is 680 pH, the capacitance is 1 pF, and the junction critical current is 1.5~$\mu$A.}
\label{fig:figure1}\end{figure}

In this work, we demonstrate the realization of a Josephson phase qubit incorporating a crystalline silicon shunt capacitor. Silicon growth has been perfected over decades by the integrated circuit industry, and commercial-grade intrinsic silicon displays an internal quality factor in excess of $10^6$ in the low-power, low-temperature regime relevant to qubit operation. Moreover, silicon-on-insulator (SOI) technology provides a path to the incorporation of a crystalline silicon membrane into a thin film parallel plate capacitor. There was a recent demonstration of high quality-factor lumped-element ${LC}$ microwave resonators incorporating such crystalline silicon membranes, with intrinsic $Q$ in excess of $2\times 10^5$\,\cite{Weber11}. Using a variant of this process, we are able to integrate the low-loss SOI capacitor with a 1~$\mu$m$^2$ scale Josephson junction and the associated control and read-out circuitary to realize a phase qubit with significantly improved energy relaxation times.

The process flow and layer stackup are shown in Fig.~\ref{fig:figure1}(a), while the phase qubit circuit layout is shown in Figs.~\ref{fig:figure1}(b-c). The devices were fabricated from commercial SOI wafers comprising a 400~$\mu$m silicon handle, 500~nm of buried SiO$_{2}$ (BOX), and a 2~$\mu$m crystalline silicon device layer. The wafer was photolithographically patterned on the back side, and Bosch reactive ion etch (RIE) and buffered oxide etch (BOE) steps were used to create a suspended silicon membrane with area  500~$\times$~600~$\mu$m$^2$. Following an HF acid dip, blanket Al thin films were sputter deposited on the front and back sides of the wafer. The front side of the wafer was patterned and chemically etched to form the base electrode of the Josephson junction, including the qubit loop and the shunt capacitor plates; flux biasing and readout resonator structures were also formed in this layer. The native aluminum oxide was removed from the base electrode layer $via$ ion milling, the junction tunnel barrier was formed by thermal oxidation at ambient temperature (with a typical exposure of 100~mT O$_2$ for 10~minutes), and the Al junction counterelectrode was deposited and chemically etched to complete the circuit.

\begin{figure}[t]
\includegraphics[width=.48\textwidth]{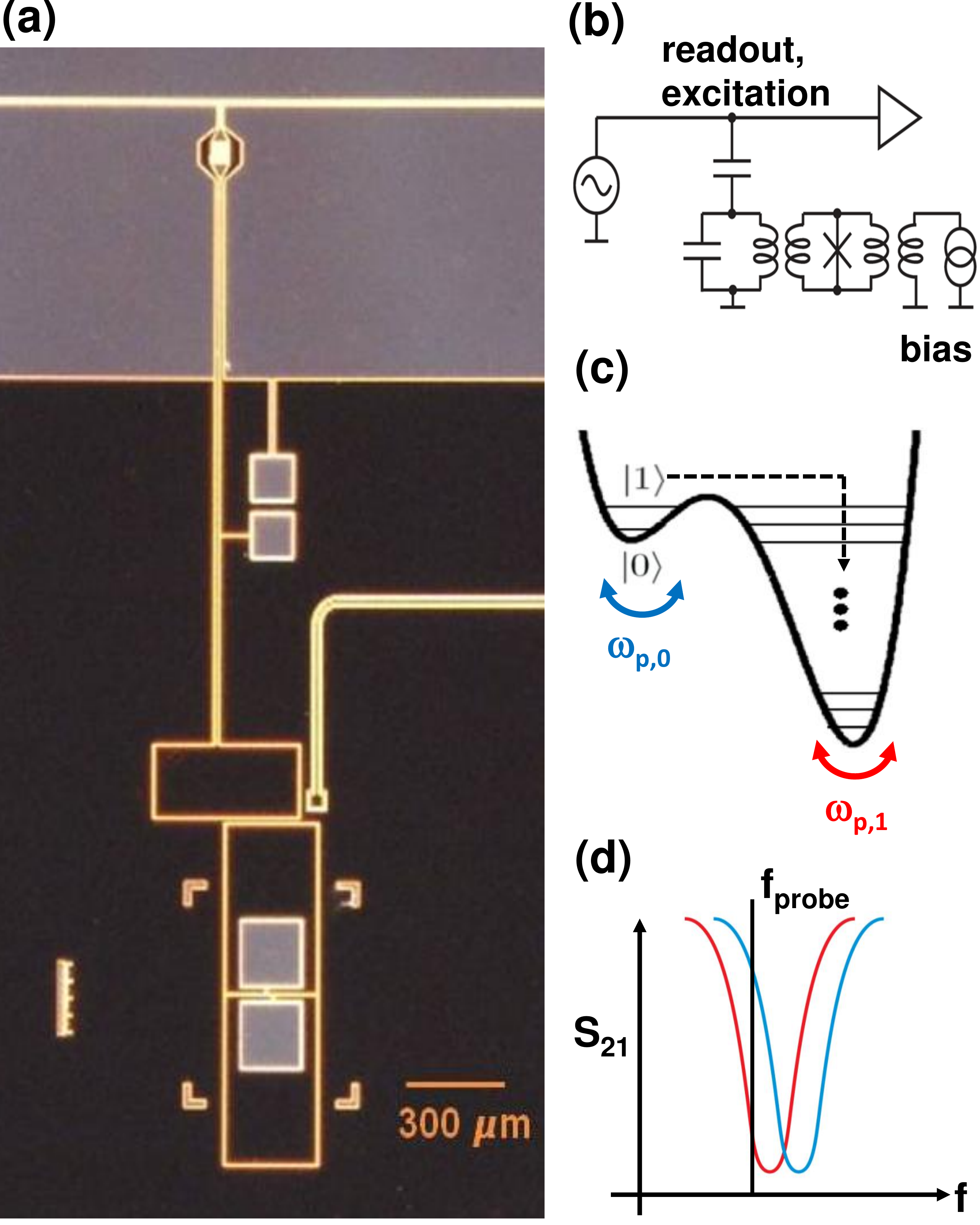}
\vspace*{-0.0in} \caption{(a) Optical micrograph of the phase qubit device. The qubit is inductively coupled to an on-chip readout resonator and flux bias coil. (b) Electronic circuit diagram of the phase qubit with biasing and readout/excitation circuitry. (c) Qubit potential energy diagram showing different plasma frequencies $\omega_{p,0}$, $\omega_{p,1}$ for the device following qubit measurement. (d) Schematic diagram of transmission across readout resonator for different qubit measurement outcomes. The probe tone is applied at a frequency that gives the largest change in transmission amplitude.}
\label{fig:figure2}\end{figure}

A micrograph of the completed device is shown in Fig.~\ref{fig:figure2}(a). The 680~pH qubit loop inductance consists of the parallel combination of two  300~$\mu$m~$\times$~500~$\mu$m Al loops with a 1.5~$\mu$A, 1~$\mu$m$^2$ Josephson junction fabricated at the middle of the central branch; this gradiometric configuration provides reduced sensitivity to fluctuations of the ambient magnetic field. The low-loss junction shunt capacitance comprises the series connection of two capacitors formed from the metallized back surface of the silicon membrane and two 200~$\times$~200~$\mu$m$^2$  plates on the top surface of the chip, with the crystalline silicon device layer acting as capacitor dielectric. The dc bias flux and fast measurement pulses are coupled to the qubit $via$ an on-chip bias coil with mutual inductance 2.4~pH to the qubit loop, and the qubit is inductively coupled to a 3~GHz $LC$ resonator used for both excitation and readout. We thus avoid the need for any explicit galvanic or capacitive connection to the phase qubit loop. The device layout is carefully symmetrized to ensure that the floating bottom electrode of the qubit capacitance acts as a virtual ground. This arrangement minimizes spurious dielectric loss in the buried oxide at the edges of the membrane cavity due to any microwave potential that might develop on the bottom electrode. We anticipate that in future work the need to symmetrize the circuit can be eliminated by explicitly grounding the bottom electrode using microfabricated vias to connect the top and bottom metallization layers through the silicon dielectric.

The 3~GHz readout resonator is capacitively coupled to a microwave feedline as shown in Fig.~\ref{fig:figure2}(b); coupling to the feedline results in a loaded $Q$ of order~1000. By driving through the readout resonator at the qubit frequency (in the range from 5-6~GHz), we are able to perform coherent qubit rotations. Qubit measurement is performed by applying a fast (few~ns) bias pulse through the flux line that projects the qubit $\ket{1}$ state out of the 01 manifold, resulting in a change in the circulating current state and effective inductance of the qubit loop. Thus, there are two distinct plasma frequencies associated with the measured qubit, as shown in Fig.~\ref{fig:figure2}(c). To read out the qubit we perform a standard homodyne measurement of the transmission across the $LC$ resonator, relying on the small shift in resonant frequency associated with the two qubit measurement outcomes as shown in Fig.~\ref{fig:figure2}(d). The qubit chip is mounted in an aluminum box and anchored at the 35~mK mixing chamber plate of a dilution refrigerator; the experiment is surrounded by a cold copper can coated with infrared-absorbing epoxy to suppress quasiparticle generation due to blackbody radiation from higher temperature stages of the cryostat \cite{Barends11}. The dc and microwave control lines are heavily filtered at 4.2~K and at the cold stage to suppress thermal excitation of the qubit.

\begin{figure}[t]
\includegraphics[width=.48\textwidth]{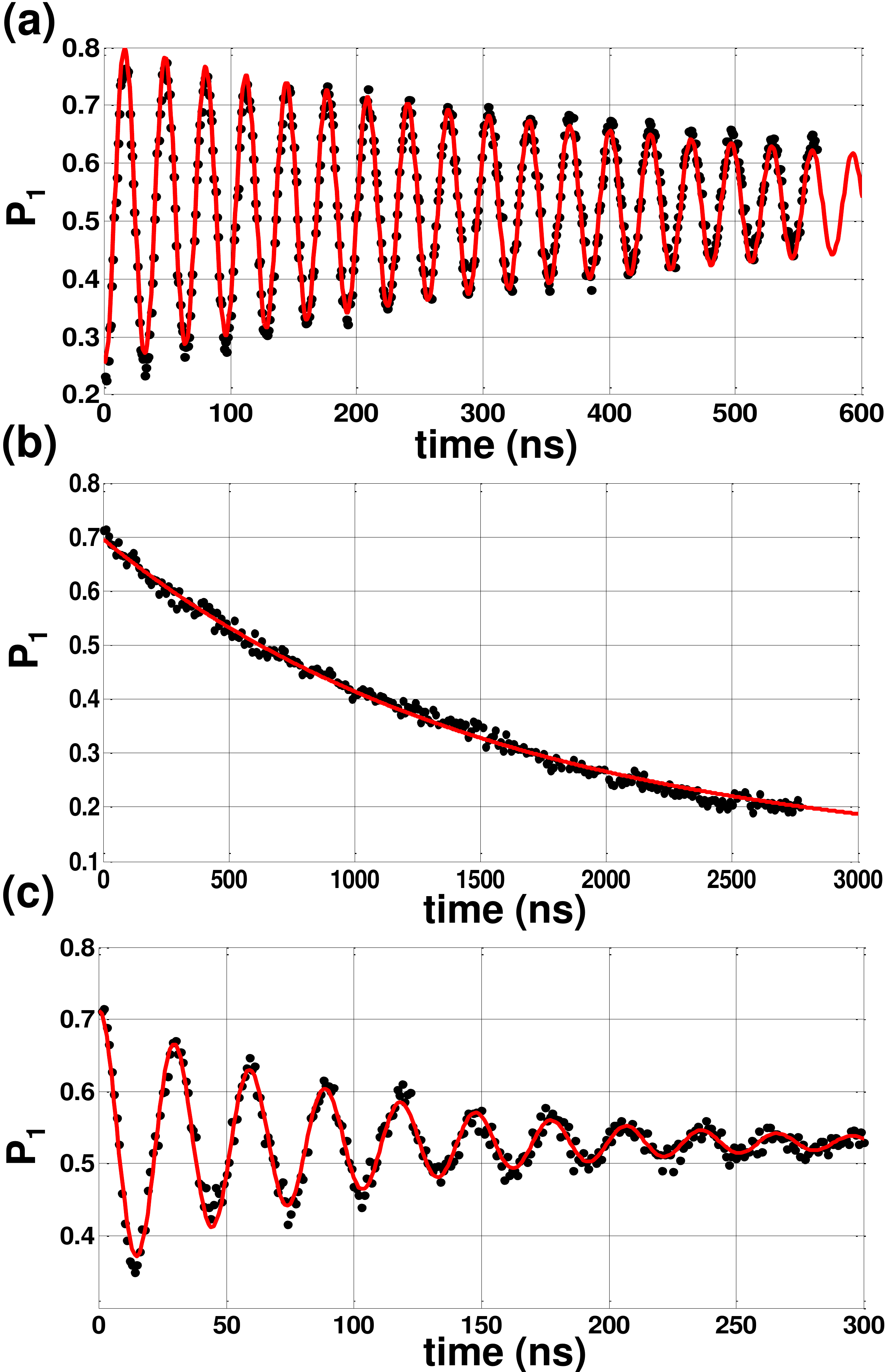}
\vspace*{-0.0in} \caption{Characterization of SOI phase qubit coherence. (a) Rabi oscillations measured at a qubit operating frequency of 5.093~GHz. The solid line is a fit to the measured curve, giving a decay time of 510~ns. (b) Spontaneous decay of the qubit excited state and corresponding fit, yielding an energy relaxation time $T_1$ = 1.6~$\mu$s. (c)  Qubit Ramsey fringes. The fit yields a dephasing time $T_2^* $ = 110~ns.}
\label{fig:figure3}\end{figure}

We have performed standard microwave pulse sequences to evaluate qubit coherence over a range of bias points; results are shown in Fig.~\ref{fig:figure3}. Qubit Rabi oscillations decay with a characteristic time in excess of 500~ns. Qubit Ramsey fringes decay on a shorter timescale of order 100~ns; as our qubit has no flux sweet spot Ramsey decay is dominated by excess low frequency magnetic flux noise \cite{Ithier05,Bialzak07,Sank12}. Qubit energy relaxation times up to 1.6~$\mu$s have been measured, and energy relaxation times greater than 1~$\mu$s have been measured on multiple devices. These SOI-based phase qubits thus display energy relaxation times that are a factor of 2-3 better than the $T_1$ times of state-of-the-art phase qubits incorporating low-loss a-Si:H shunt capacitors \cite{Wang08,Neeley08}. It is worthwhile to consider the coherence budget of our qubit device. First, dielectric loss in the crystalline silicon shunt capacitor presents an ultimate limit to qubit $T_1$. However, the measured loss tangent of $5\times 10^{-6}$ would correspond in our device to an energy relaxation time around 10~$\mu$s, far in excess of the measured $T_1$. Next, qubit $T_1$ is limited by coupling to the heavily damped readout resonator, a phenomenon known as the Purcell effect \cite{Houck08}. However for our device and operating parameters the Purcell limit to qubit $T_1$ is around 20~$\mu$s. Instead the current devices appear to be limited by overcoupling to the flux bias coil: experiments reveal an oscillatory dependence of $T_1$ on qubit bias frequency which we have determined to arise from a frequency-dependent impedance transformation of the dissipation looking into the flux line. We anticipate that it will be straightforward to suppress this decoherence channel by re-engineering the bias tee on the flux line and by reducing the mutual inductance from the flux bias coil to the qubit.

\begin{figure}[t]
\includegraphics[width=.48\textwidth]{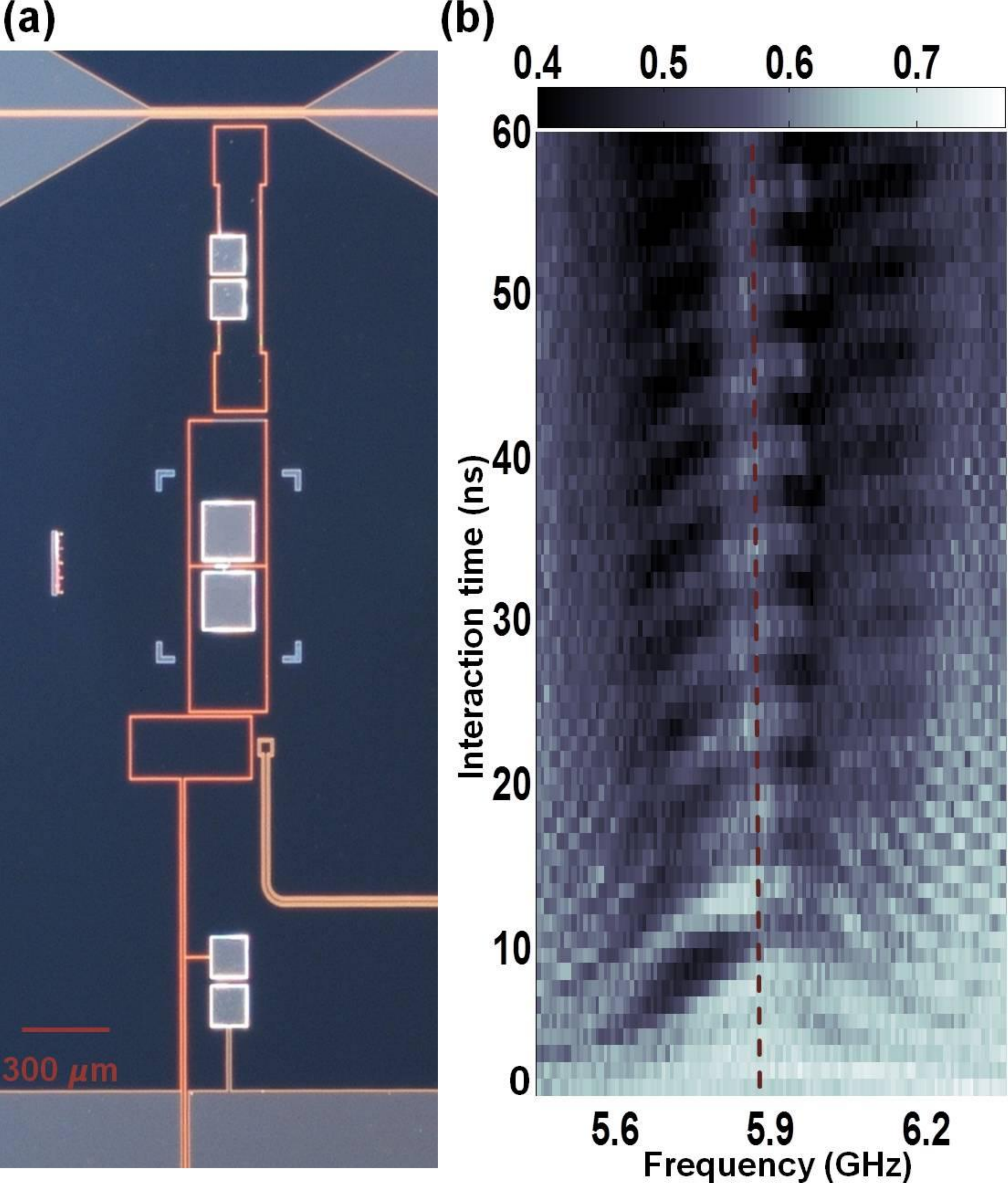}
\vspace*{-0.0in} \caption{Swap spectroscopy for qubit-$LC$ resonator system. (a) Optical micrograph of circuit with one SOI phase qubit coupled to two SOI-based microwave $LC$ resonators. (b) Qubit excited state occupation $P_1$ \textit{versus} operating frequency and qubit-resonator interaction time. Coherent interaction between the qubit and the resonator yields a characteristic chevron pattern with a swap frequency of 190~MHz on resonance (dashed line).}
\label{fig:figure4}\end{figure}

The SOI-based phase qubit can be adapted to realize more complex quantum circuits without any additional fabrication steps. We have fabricated a circuit in which one SOI phase qubit is inductively coupled to two on-chip $LC$ resonators; a micrograph of this circuit is shown in Fig.~\ref{fig:figure4}(a). With this circuit we we have performed swap spectroscopy of the $LC$ resonators \cite{Marian11a,Marian11b}. In this experiment, the qubit is initially detuned from the resonator and a microwave $\pi$-pulse is used to promote the qubit to the excited state. Next, a fast bias pulse is used to bring the qubit into or close to resonance with the $LC$ mode, and the two systems are allowed to interact for a variable period of time. Finally, the qubit is measured in the conventional way. In Fig.~\ref{fig:figure4}(b) we show swap spectroscopy of an $LC$ mode at 5.89~GHz. The false color plot of qubit excited state occupation $P_1$ \textit{versus} frequency and interaction time displays a characteristic chevron pattern, with coherent swaps on resonance at the coupling frequency 190~MHz.

In summary, we have characterized the coherence of phase qubit circuits incorporating single crystal silicon capacitors derived from SOI wafers. The qubit $T_1$ times are more than a factor of two greater than those observed in state-of-the-art phase qubits incorporating low-loss amorphous shunt capacitors. The SOI phase qubits are readily integrated into more complex circuits incorporating multiple qubits and resonators. Thus SOI technology could provide a path to scalable, coherent multi-qubit circuits for rigorous investigations of quantum algorithms.\\

\begin{acknowledgments}
This work was supported in part by IARPA through grant W911NF-09-1-0368 and by the National Science Foundation through grant DMR-0805051. Some work was performed at the Cornell NanoScale Facility, a member of the National Nanotechnology Infrastructure Network, which is supported by the National Science Foundation (Grant ECS-0335765).
\end{acknowledgments}

\end{document}